\documentstyle[aps,epsfig]{revtex}
\begin{document}

\newcommand{\beq}{\begin{equation}}
\newcommand{\eeq}{\end{equation}}
\newcommand{\A}{\mbox{\bf A}}
\newcommand{\Ss}{\mbox{\bf S}}
\newcommand{\Ha}{\mbox{\cal H}}

\title{Statistical Properties of Contact Maps }
\author{
Michele Vendruscolo$^{(1)}$, 
Balakrishna Subramanian$^{(2)}$,
Ido Kanter$^{(3)}$,
Eytan Domany$^{(1)}$ and
Joel Lebowitz$^{(2)}$}
\address{$^{(1)}$Department of Physics of Complex Systems, Weizmann Institute of Science, Rehovot 76100, Israel \\
$^{(2)}$Department of Mathematics, Rutgers University, New Brunswick NJ 08903 \\
$^{(3)}$Department of Physics, Bar Ilan University, 52900 Ramat Gan, Israel}

\address{
\centering{
\medskip\em
{}~\\
\begin{minipage}{14cm}
{}~~~
A contact map is a simple representation of the structure of proteins and
other chain-like macromolecules. This representation is  quite  
amenable to numerical studies of folding.
We show that the number of contact maps corresponding to the possible
configurations of a polypeptide chain of $N$ amino acids,
represented by $(N-1)$-step self avoiding walks on a lattice,  
grows exponentially with $N$ for all dimensions $D > 1$.
 We carry out exact enumerations in $D=2$ on the square and 
triangular lattices for walks of up to 20 steps and investigate 
various statistical properties of contact maps corresponding 
to such walks.. We also study the exact statistics of contact maps 
generated  by walks on a ladder. 
{}~\\
{}~
\end{minipage}
}}
\maketitle

\section{Introduction}

Prediction of a protein's structure from its amino acid sequence
is an important and  challenging open problem.
The first choice one has to make when approaching the problem is that 
of {\it structure representation}. 
One of the most minimalist representations of a protein's structure is in 
terms of its {\it contact map} \cite{ls79,hck79} which, for
a polypeptide chain of length $N-1$, 
is an $N\times N$ matrix \mbox{\bf S}. Denoting by $i,j$ 
the position index of the amino acids 
along the chain, the elements of \mbox{\bf S}
are defined as 
\beq
S_{ij}=
\left\{
\begin{array}{ll}
1 \qquad  & \mbox{\rm if amino acids $i$ and $j$
are in contact} \\ 
0 & {\rm otherwise}
\end{array} \right.
\eeq
``Contact'' can be defined in various ways: for example
\cite{md96}, one can set $S_{ij}=1$ when there exist two
heavy (all but hydrogen) atoms, one from amino
acid $i$ and one from $j$, separated by  less than some threshold distance.
Contact maps are independent of the 
coordinate frame used 
and for compact structures, such as  the native state of proteins, 
with
many contacts, it is relatively easy to go from a map to a set 
of possible structures to which it may correspond \cite{hck79,ch88,vkd97}. 
On a lattice, a protein conformation, or fold,
is represented as a self-avoiding random walk (SAW) \cite{ld89}.
A site on the lattice visited by the walk  represents
an amino acid.
Two sites of the SAW are in contact if they are nearest neighbors
and they are {\it non-consecutive} along the walk.

To  search for a protein's native structure in 
the space of contact maps 
(as has been proposed by several groups),  it is important
to have general knowledge about the size and nature of this space.
Recent studies \cite{ib1,ib2} of the dynamics of naturally occurring proteins 
has shown that the contact maps along with simple energetics 
is enough information   
to reproduce the vibrational spectrum with some  
accuracy. This makes it important to understand the
statistics of the contact map representation. 
In particular, one would like to know how the number of different 
{\it physical} 
contact maps depends on the chain length $N$. 
Clearly one has $2^{N(N+1)/2}$ distinct $N \times N$ symmetric 
matrices of binary elements $S_{ij}=0,1$. 
Most of these, however, do not correspond to physical structures; 
these  matrices 
cannot be realized as contact maps  of
real, physical chains or SAW's. 

In fact, as we shall see, $N_M$,
the total number of physical maps obtainable for 
a chain of length $N$ on a lattice satisfies the bounds 
\beq
e^{c_\ell N} \leq N_M \leq e^{c_u N} \; .
\label{eq:expon}
\eeq
The upper bound in (\ref{eq:expon}) follows trivially from the bound
on $N_{SAW}$, the total number of SAW's, which is asymptotically
given by \cite{saw}
\begin{equation}
N_M \le N_{SAW} \sim N^{\gamma-1} \mu^N
\sim e^{c_u N} \; , c_u = \ln{\mu},
\label{eq:nsaw}
\end{equation}
where $\mu$ is the connectivity constant of the lattice, and $\gamma$
is a critical exponent.

A simple construction of a special set of walks, each with a distinct contact
 map  provides the lower bound. Start the chain at the
the origin, $i=1$; the first step and all odd-indexed steps are in the
positive horizontal direction $+x$, whereas every even indexed step is in the
vertical direction, either $+y$ or $-y$. The decision taken for step
$2i$ either brings the site $2i+1$ into contact with $2i-2$, in which case
$S_{2i-2,2i+1}=1$, or this contact is absent and $S_{2i-2,2i+1}=0$. Hence
for every choice of the set of vertical steps we get a walk whose contact
map does not appear for any different walk from the set. Clearly, the maps
constructed this way must have $S_{2i-k,2i+1}=0$ for $k>1$.
In this way we obtain $N_{SAW}^{\prime}$ SAW's
and the following exponential lower bound for the number of contact maps
\begin{equation}
N_M \ge N_{SAW}^{\prime} \sim 2^{N/2} \sim e^{c_\ell N} \; .
\label{eq:lower}
\end{equation}
Clearly the argument works for any dimension and can be extended for 
the triangular lattice. This (rather poor)
bound can be improved by including walks whose maps can have other
non-vanishing elements, e.g. with $S_{2i-4,2i+1}=1$.
A better lower bound for the square lattice is obtained by an explicit
 construction given in Section 3.

We actually expect that $\ln(N_M)/N$ approaches a limit,
\beq
\frac{\ln(N_M)}{N} \rightarrow a \;
\label{eq:expon1}
\eeq
as N becomes large (the existence of a limit does not follow directly
from (\ref{eq:expon})). 
To estimate $a$ we  
computed, for $N \leq 20$, the precise numbers $N_M$ 
on the square and triangular lattices. 	
This is done by exact enumeration of all possible distinct SAW's,
i.e. not related by symmetry operations of the lattice,
and the corresponding contact maps. 

Using this enumeration for low $N$ and  sampling 
for larger $N$, we also computed various other statistics of
contact maps, such as the number of maps with  particular density
of contacts,
the number of SAW's that correspond to this set of maps, etc.
We also obtained explicit results about the corresponding 
quantities for walks on a special ``ladder'' lattice.

\vspace{.5truecm}

\section{Exact enumeration in $D=2$}
\begin{figure}
\centerline{\epsfig{figure=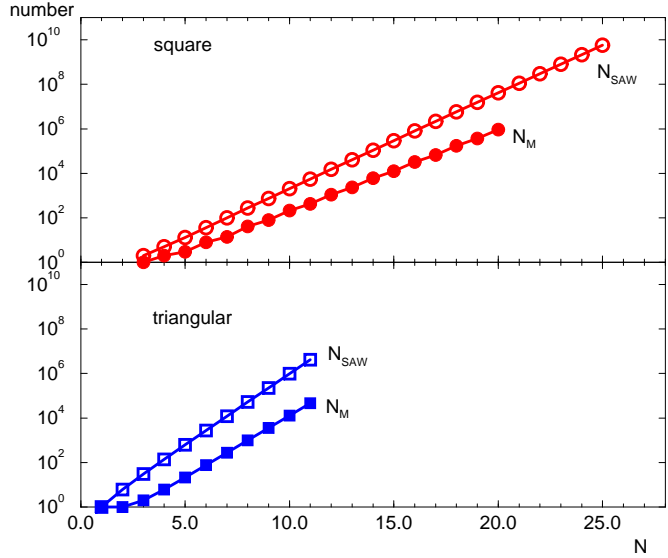,height=7.5cm,angle=0}}
\caption{
Upper curve: $N_{SAW}$, the number of SAW's versus their
length $N$, obtained by exact enumeration for $N \le 25$
on the $2D$ square lattice and fitted with Eq. (\protect\ref{eq:expon}).
The lower curve shows the exponential
variation of $N_M$,  the number  of contact maps
corresponding to all possible SAW's with $N \le 20$.
Data were obtained from complete enumeration.
}
\label{fig:nms}
\end{figure}

\begin{figure}
\centerline{\epsfig{figure=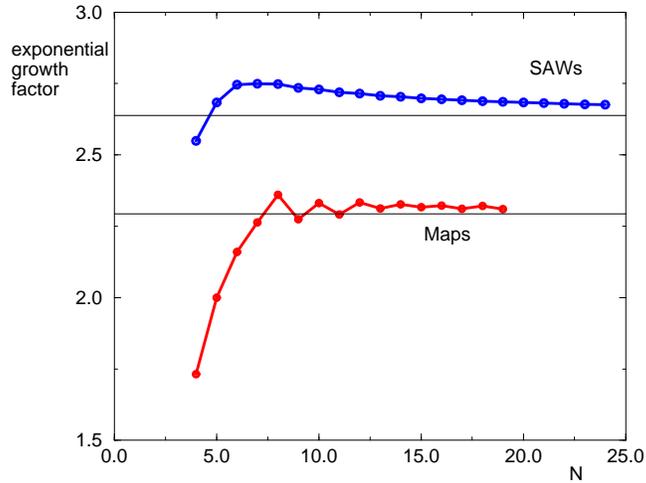 ,height=7.5cm,angle=0}}
\caption{A comparison of the connective constant $\mu$ for the walks and
the exponential growth factor $\exp{(a)}$ for the contact maps generated
on a square lattice evolving with the size of the walk.
Horizontal lines are the known value $\mu=2.638$ \protect\cite{saw}
and our estimate $e^a=2.3$, based on the data for all $N$'s.}
\label{fig:slope}
\end{figure}

In the upper curve of Fig. \ref{fig:nms} we plot
the number of walks $N_{SAW}$,
obtained by complete enumeration \cite{rapaport},
versus $N$, fitted (for $N \leq 25$)
with the known \cite{saw} estimates $\mu=2.6381585$ for the connective constant
and $\gamma = 43/32$ for the critical exponent on the $2D$ square lattice.
The lower curve is the total number $N_M$ of contact maps,
corresponding to all possible SAW's with
$N \le 20$ on a $2D$ square lattice.
Fitting Eq.(\ref{eq:expon}), we obtain $a=0.83(1)$.
This result was obtained previously, by enumerating walks with $N \leq 14$,
by Chan and Dill\cite{CD91}.
For comparison, we note that a straightforward fit of $N_{SAW}$
with Eq.(\ref{eq:nsaw}) gives the upper bound prefactor $c_u =1.00(1)$,
and that the lower bound prefactor, as from Eq.(\ref{eq:lower}),
is $c_{\ell}=0.346$.
We obtained the corresponding results for the triangular lattice.
But in this case due to the higher density of contacts,
we were able to obtain results
only for $N \leq 11$ as shown in Fig. \ref{fig:nms}.
Our fit gave $c_u =1.47(1)$ and $a=1.28(1)$.
To address the  question  whether in $D=2$ the constant $a$ for
the contact maps is strictly less than $\ln(\mu)$
, we present in
Fig. \ref{fig:slope} a plot of the running value of the connective 
constant $\mu$ for the walks and the running value of $\exp{(a)} $
for the maps as the size of the walk increases. 
The running slope $\mu(N)$ is computed from 
enumeration data using the standard formula
\begin{equation}
 \ln \mu(N) = \frac{1}{2}\left [ \ln N_{SAW}(N+1) - \ln N_{SAW}(N-1) \right ]\; ,
\end{equation}
and an analogous one for the factor $a$ for maps.

This figure is consistent with $a < \ln \mu$.

Most biologically functional proteins  fold into remarkably compact
conformations, with very few solvent molecules in the interior.
Therefore it is of interest to consider
how the number of contact maps
and their corresponding walks varies  with the number of contacts.

\begin{figure}
\centerline{\epsfig{figure=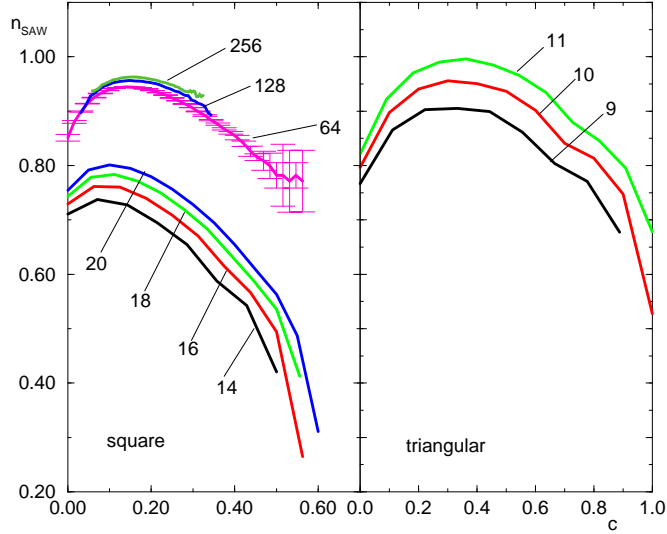,height=7.5cm,angle=0}}
\caption{
Logarithm $n_{SAW}(c)$ of the fraction of 
walks with a given fraction $c$ of contacts.
On the square lattice, we show data obtained from exact enumeration
for chain lengths $N=14,16,18,20$,
and data obtained from sampling for $N=64,128,256$.
For clarity, errors on data from sampling are shown only for $N=64$.
On the triangular lattice, we show the 
fraction  $n_{SAW}(c)$ of walks with a given number $Nc$ of contact
for $N=9,10,11$.
}
\label{fig:nwnc}
\end{figure}

Denote  by
$N_{SAW}(c)$ the number of walks with a fixed number $Nc$ of contacts. When
 there is an interaction energy $u$ associated with each contact, then 
$\ln(N_{SAW})$ is identical to the entropy  of the chain at energy $E= Ncu$. 
In Fig. \ref{fig:nwnc} we plot the fractions 
\begin{equation}
n_{SAW}(c) = \ln(N_{SAW}(c)/N)
\end{equation}
for chains of different lengths $N$ on the $2D$ square lattice.

The time required to enumerate walks and maps
increases exponentially with the size $N$ 
and it becomes impractical to use this method.
However we want to generate the statistics for larger values of N, which
is the actual physical situation.
Standard techniques are routinely used \cite{saw}  
to generate unbiased samples of SAWs on the lattice. 
We use the method of incomplete enumeration (Redner-Reynolds) 
to generate our sample of unbiased SAW's.

We use our sample to generate the distribution of the fraction of
the walks with a given number of contacts $n_{SAW}(Nc)$ 
as introduced before for SAW's 
of length N=64,128,256. 
In Fig. \ref{fig:nwnc}, we plot the result.

\begin{figure}
\centerline{\epsfig{figure=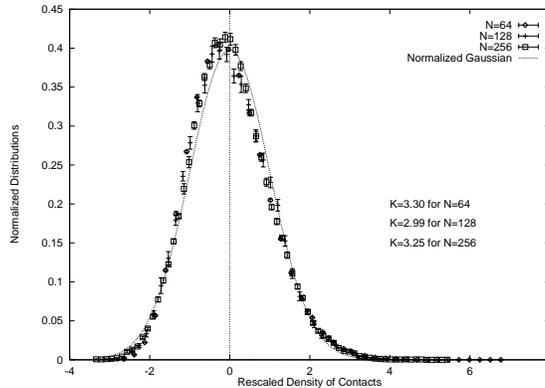,height=7.5cm,angle=-90}}
\caption{
The collapse of distributions(for three different lengths) for the fraction
of walks with a given number of contacts after rescaling the finite-size
variables.
}
\label{fig:scal}
\end{figure}

One would like to say something about how
this distribution looks in the asymptotic limit. We try to analyze 
this by rescaling the finite-size variables such that the distributions 
collapse on top of each other. If Fig. \ref{fig:scal}, we observe 
that normalizing the variance to 1 and the mean to 0, results in the 
collapse of the distributions (for the three different lengths of N=64,
128,256). We compare this to the normalized Gaussian. From the data
obtained, it appears that we cannot rule out either possibility (Gaussian or 
non-Gaussian). We also list the kurtosis values K obtained for the different
data sets. For a Gaussian distribution, we expect an exact value of 3.00. 


It is however not clear how one should generate the distribution
of the maps with a given number
of contacts  $N_M(Nc)$. While we have standard and efficient algorithms
to generate SAWs with the desired weight, it seems difficult
to generate contact maps which are equally 
weighted in the sample and not biased with their degeneracies.

\begin{figure}
\centerline{\epsfig{figure=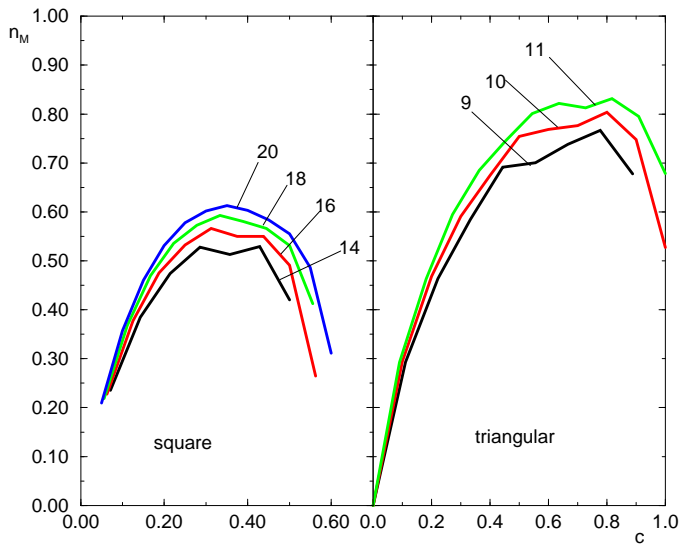,height=7.5cm,angle=0}}
\caption{
Logarithm $n_M(c)$ of the fraction 
of contact maps with a given fraction $c$ of contacts,
shown for 4 different walk lengths on the 2D square lattice, and 
 for 3 different walk lengths on the triangular lattice.
}
\label{fig:nmnc}
\end{figure}

Let now $N_M(c)$ be the number of distinct contact maps with $Nc$ contacts.
We show in Fig. \ref{fig:nmnc} how the fractions 
\begin{equation}
n_M(c) = \ln \left(N_M(c)/N\right) 
\end{equation}
vary with $c$, again for various chain lengths.

The main difference between Figs. \ref{fig:nwnc} and \ref{fig:nmnc}
is that the distribution of walks has its maximum at smaller
values of $c$ than the distribution of contact maps. This can be
understood intuitively by a noting that  for small $c$ the number of maps is
suppressed in comparison to  $N_{SAW}(c)$:
for example, there is only a single contact map with $c=0$,
whereas there are many walks with no contacts.
In general, the degeneracy of contact maps
has a non-trivial dependence on the number of their contacts.

\begin{figure}
\centerline{\epsfig{figure=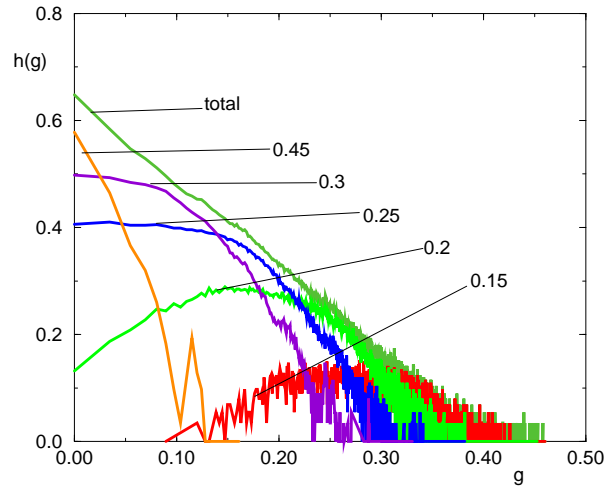,height=7.5cm,angle=0}}
\caption{
Histogram of $h(g)=\ln H(g)/N$, where $H(g)$ is the number of
maps with degeneracy $G=e^{Ng}$, 
for walks of length $N=20$ on the square lattice.
Separate curves are shown for subset of maps with $c=0.15,0.2,0.25,0.3,0.45$.
}
\label{fig:hg}
\end{figure}

Consider walks of length $N$ and denote by
\begin{equation}
G=e^{Ng}
\end{equation}
the degeneracy of a map, i.e.
the number of walks corresponding to that contact map.
For each value of $g$ we determined $H(g)$,
the number of maps whose degeneracy is $e^{Ng}$.
This information is shown in Fig. \ref{fig:hg}
where we present $h(g)=\ln H(g)/N$ versus $g$,
for walks of length $N=20$ on the square lattice.
We further analyze the degeneracy by concentrating
on the {\it subset of maps} with a fixed number of contacts, $Nc$.
In Fig. \ref{fig:hg} we show results for $Nc=3,4,5,6,9$, 
i.e. $c=0.15,0.2,0.25,0.3,0.45$.
Not surprisingly, the maps with large number of contacts which 
correspond to the typical native folds of proteins generally have small
degeneracy. It is the maps with few contacts which account for 
the large degeneracy. In general the map with $c=0$ (all zero entries in 
the matrix) has $G > 2^N$ corresponding 
to all the directed walks with no contacts. 
The walks that correspond to
maps with different degeneracies differ in the lengths of
contact-free segments that the walk has.
For $N=20$ and $Nc=6$ on the square lattice, we measured the length $L$ of the
longest contact-free stretch at the ends of the walk.
Maps with low degeneracy have, on the average, $L \simeq 1$, whereas for
highly degenerate maps we found, typically, $L \simeq 7$ (there
are also highly degenerate maps and walks with 
long contact-free stretches far from the ends).
Clearly, the presence of long stretches free of contacts is responsible
for the high degeneracy of a map.

\begin{figure}
\centerline{\epsfig{figure=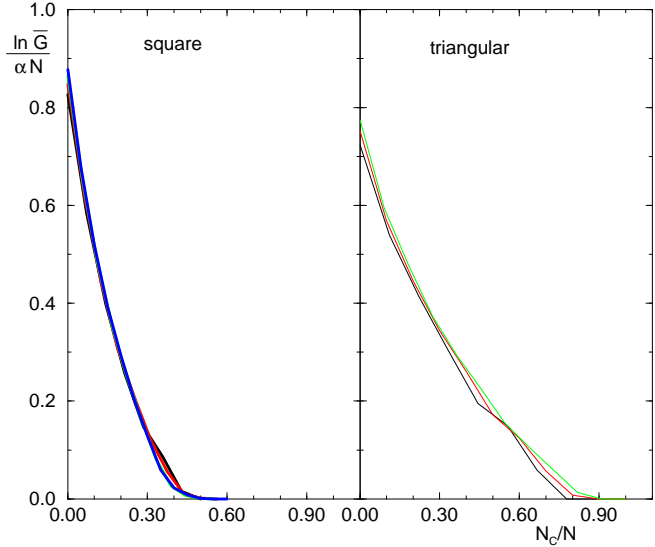,height=7.5cm,angle=0}}
\caption{
Scaling plot of the degeneracy function $\bar{G}(c)$
averaged over all the contact maps
with $Nc$ contacts, plotted for
different chain lengths $N$ on the triangular and square lattices.
}
\label{fig:hag.sca}
\end{figure}

Let now $\bar{G}(Nc)$ denote the average degeneracy over all the
maps with $Nc$ contacts.
We studied $\bar{G}(Nc)$ as a function of $Nc$.
As already mentioned, contact maps corresponding to maximally compact walks have, on the average,
a very small degeneracy.
It seems reasonable to assume that for a fixed $ c = \frac{N_c}{N}$, 
$\bar{G} (Nc)$ will grow exponentially with N, such that

\begin{equation}
\ln \bar{G}(N,Nc) = \alpha N f\left (c \right ) \; .
\label{eq:logG}
\end{equation}
The enumeration results seem to support this assumption as seen in
the collapse plot in Fig. \ref{fig:hag.sca}
with $\alpha =0.86$ for the square lattice
and $\alpha =1.07$ for the triangular lattice.
The value of $\alpha $ is extracted by fitting
$\bar{G}(N,0) \sim e^{\alpha N}$.
As we can see the assumption Eq. (\ref{eq:logG})  
seems to hold to good accuracy.

\section{Exact Results for walks and maps on a ladder}
In this section, we introduce and solve the problem exactly on a 
toy lattice. 
The lattice is a ladder of two rows of sites, at points $(x,y)$ with  $y=0,1$
and $x=0,1,2,....$. We consider all walks starting at the origin
with  horizontal steps in the positive $x$ direction.
We first show that the numbers of SAW's and contact maps
is exponential in $N$, with different coefficients $a$. 
Denote by $A(N)$ the number of walks of $N$ steps; 
\[
A(N)=A_h(N)+A_v(N)
\]
where $A_h(N)$ is the number of walks that 
end with a horizontal step and  $A_v(N)$ walks end with a
vertical step. 
Since a vertical step {\it must} be preceded by a horizontal one we have
\[
A_v(N)= A_h(N-1)
\]
On the other hand, to every walk one can add a horizontal step so that
\[
A_h(N)=A(N-1)
\]
Thus we get, using these three relationships, the recursion for the Fibonacci
numbers:
\begin{equation}
A(N)=A(N-1)+A(N-2)
\label{eq:Fib}
\end{equation}
and hence the number of walks grows, for large $N$, exponentially
\begin{equation}
A(N) \sim e^{a_w N} \qquad \qquad a_w = {\rm ln} \frac{1+\sqrt{5}}{2} \approx
0.481
\label{eq:check}
\end{equation}
A recursion for the number of contact maps can be calculated as well. 
One way to do
this is by representing $B(N)$, 
the total number of distinct contact maps of $N$
steps as a sum
\[
B(N)=B_0(N)+B_1(N)
\]
where $B_0(N)$ is the number of contact matrices (maps) whose first row
contains only zeroes (i.e. the first site does not have a contact); $B_1(N)$
is the number of those maps for which the first site does have a contact.
Since to every map we can add a first row (and column) of zeroes, we have
\[
B_0(N)=B(N-1)
\]
For all maps that start with a contact, the first four steps are fixed;
the corresponding walks can be continued in two different ways, either with
a vertical step or with a horizontal one. These two possibilities give rise to
a recursion of the form
\[
B_1(N)=B_1(N-2)+B(N-5)
\]
With a little algebra the last three equations yield
the final recursion
\begin{equation}
B(N)=B(N-1)+B(N-2)-B(N-3)+B(N-5)
\label{eq:Maps}
\end{equation}
If we now assume that 
\[
\frac{\ln(B(N))}{N} \rightarrow e^{a_m}
\]
as N becomes large, we find that $e^{a_m}$ is the solution  of the equation
\[
q^5-q^4-q^3+q^2-1 =0
\]
which yields
\[
a_m \approx 0.367 < a_w
\]
Having counted the number of walks and maps, we turn to
calculate various
statistical features 
of maps and walks on a ladder. 
For example, we can consider the fraction of maps with a given  number of 
contacts; 
the degeneracy of maps, i.e. the number of different walks that have the same
map,etc. Analytical examination of such  quantities  sheds light on
the origin of results obtained from exact enumeration of walks
in two dimensions and indicate the extent to which the relatively short
chains that can be enumerated represent the true two dimensional
behavior.

\begin{figure}[h]
\centerline{\psfig{figure=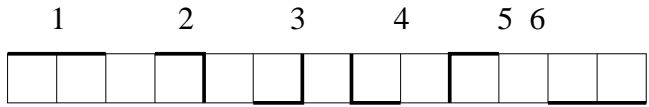,height=0.5in}}
\caption {The six possible two successive steps on a ladder.
	}
\label{fig:2steps}
\end{figure}

A walk of $N$ steps taken according to the rules given above can be 
characterized by the sequence of the contact-free intervals between all pairs
of consecutive contacts. We denote by $m$ the number of steps needed to walk
from the end site of contact $n-1$ to the end site of contact $n$.
Let $D(m)$ denote the degeneracy of such a contact-free walk, i.e. the number 
of different SAWs of length $m$. To calculate 
$D(m)$, we introduce a transition matrix $L$, among six possible ``states'',
of pairs of consecutive steps, referred to as ``2-steps''. 
The six possible 2-steps that can occur on a 
ladder
are shown in Fig. \ref{fig:2steps}. 
The fact that $L_{42}=1$ shows that 
it is possible to have a 3-step walk whose first and second steps constitute a
2-step is of type 2 and	the second and third step constitute a 2-step of type 4.
Note that only those transitions that do not terminate the walk (i.e. do not 
generate a contact) are designated as possible by the matrix $L$ - for example
we have $L_{34}=0$ since a 4 followed by a 3 generates a contact.


\begin{equation}
L= \left(
\begin{array}{cccccc}
1  &  0  &  0  &  0  &  1  &  0 \\
1  &  0  &  0  &  0  &  0  &  0 \\
0  &  0  &  0  &  0  &  0  &  1 \\
0  &  1  &  0  &  0  &  0  &  0 \\
0  &  0  &  1  &  0  &  1  &  0 \\
0  &  0  &  0  &  1  &  0  &  1 
\end{array} \right)
\end{equation}
Note that to have a contact by adding 2-step $n$, the 2-step $n-1$
must be either of type 4 or 5; the corresponding 
vectors are $V_4 = (000100)$ and $V_5 = (000010)$. The degeneracy  of walks
of length $m$ in between contacts is then given by
\begin{equation}
D(m)= \lbrack (V_4)^T + (V_5)^T \rbrack  L^{m-1} V_2 ~~~or~~~
\lbrack (V_4)^T + (V_5)^T \rbrack  L^{m-1} V_3
\label{eq:Dm}
\end{equation}
\noindent The possible lengths for  $m$ are $2,~5,~6,~7,~ . . .$, and the 
corresponding degeneracies are given $1,1,1,1,2,3,4,6,9,13,19,28,41,60...$.
Note that $D(100) \sim 10^{16}$ and asymptotically 
\begin{equation}
D(m) \propto (1.465)^m = e^{0.382m}
\end{equation} 
\noindent 
where  $1.465$ is the largest eigenvalue of the matrix $L$.

An $N \times N$ contact map is completely specified by the set of inter-contact
intervals $\{ m \}$. If for a given map an interval of
length  $m$ appears $N(m)$ times, denote 
\[
P(m)=N(m)/N
\]
The 
logarithm of the number 
of SAW associated  with this particular map is then given by 
\begin{equation}
\ln N_{SAW}(\{ m \} ) =
N \sum_m P(m) \ln D(m) 
\end{equation} 
\noindent 
The number of contacts of this map is given by 
\begin{equation}
N_c( \{ m \} ) = \sum_m N(m) = N \sum_m P(m) = N c
\end{equation}
where the number of contacts per step,
\begin{equation}
c= \sum_m P(m)=\frac{N_c}{N}
\end{equation}
\noindent
The normalization of the $P(m)$ is 
\begin{equation}
\sum_m P(m) m = 1
\end{equation} 
\noindent The number of maps, $N_M$,  characterized by the same set of 
fractions $\{P(m)\}$
(with different orderings of the contact-free intervals) is
\begin{equation}
\ln N_M = - N \sum_m P(m) \ln P(m)  + N c \ln c
\end{equation}  
\noindent and therefore the number of SAW, $N_{W}$, associated with all
maps characterized by the same fractions $\{P(m)\}$ is
\begin{equation}
\ln N_{SAW}({P(m)}) = - N  \left\{ \sum_m  P(m) [  \ln P(m) - \ln D(m)] 
-c\ln c \right\} 
\end{equation}  

The interplay between these  two terms is  clear. As
the distance $m$  between contacts increases, the number of SAW
corresponding to such a map increases exponentially, but at the same time the 
number of contacts in the  map decreases and the number of such maps 
(permutation of the distances) decreases exponentially.
Some limiting cases can be analyzed as follows. 
For the case densest with contacts, i.e. 
$c=0.5$,  there are   only two  possible
maps  and  hence $\ln( N_{SAW})/N \rightarrow 0$.  On the other hand,  
for maps  with $O(1)$ contacts, and hence $c \rightarrow 0$, 
$m$ scales with $N$ and  $D(N) \propto e^{0.382N}$,
and therefore $N_{SAW} \sim e^{0.382N}$. 
Since in  both limiting cases  
$ \ln(N_{SAW})$  does not scale as  $0.481N$ 
(see eq. (\ref{eq:check})), the quantity $ \ln (N_{SAW})$ is expected to have 
a maximum at some intermediate value of $c$.  

The number of SAW associated 
with maps that have 
$Nc$ contacts can be studied analytically;  
\begin{equation}
N_{SAW}(c)=
\int_0^1  \pi dP(m) \delta \left[ \sum_m mP(m)-1 \right] \delta
\left[ \sum_mP(m)-c \right]
e^{-N\left\{ \sum_m  P(m)[ \ln P(m) -\ln D(m)]
-c\ln c \right\} }
\label{eq:NSAW}
\end{equation}

\begin{figure}[h]
\centerline{\psfig{figure=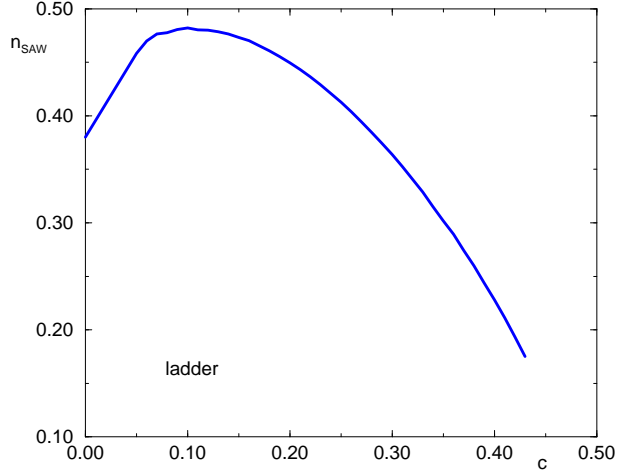,height=7.5cm,angle=0}}
\caption {$n_{SAW}=\ln \left( N_{SAW}/N \right)$ 
versus $c$, for maps of $Nc$ contacts}
\label{nwnc.ladder}
\end{figure}

\noindent 
The integrals are evaluated by the saddle point method; the
resulting equations can be reduced to the following 
coupled equations for $P(2)$ and $P(5)$ 
\begin{eqnarray}
1 = P(2)\sum_m D(m) \lbrack P(5)/P(2) \rbrack^{(m-2)/3}m \nonumber  \\
c = P(2)\sum_m D(m) \lbrack P(5)/P(2) \rbrack^{(m-2)/3} 
\label{eq:saddle1}
\end{eqnarray}
\noindent where for every allowed $m=2,5,6,7,....$, 
the degeneracy $D(m)$  is determined by eq. (\ref{eq:Dm}); these are 
supplemented by
\begin{equation} 
P(m)=P(2) D(m) \lbrack P(5)/P(2) \rbrack^{(m-2)/3}
\label{eq:saddle2}
\end{equation}
The numerical solution of the saddle  point equations gives 
$\frac{1}{N}\ln N_{SAW}$ 
as a function of $c$, the concentration of 
contacts, is presented in Fig. \ref{nwnc.ladder}. 
The maximum $\frac{1}{N}\ln(N_{SAW})=0.481$, as expected, 
is obtained for $c\sim 0.105$.

One can calculate $\ln (N_{M})$ as a function of $c$ in a similar fashion.
All one has to do is to  set $D(m)=1$ in eq. (\ref{eq:NSAW}); the 
resulting saddle point equations are obtained from eq. (\ref{eq:saddle1}-
\ref{eq:saddle2}), by using there, again, $D(m)=1$.

\begin{figure}[h]
\centerline{\psfig{figure=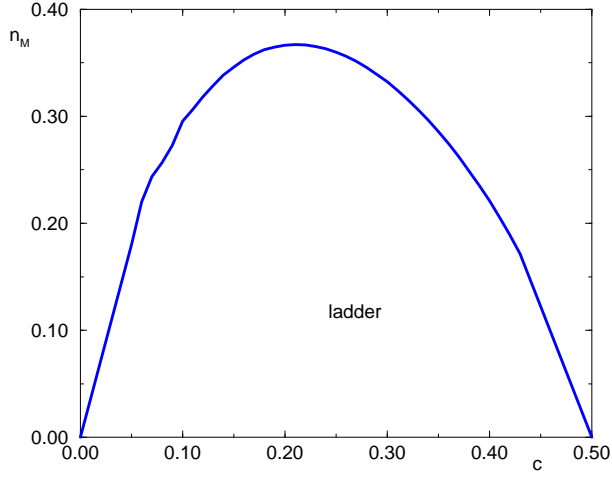,height=7.5cm,angle=0}}
\caption {$n_M= \ln \left(N_{M}/N\right)$ versus $c$;  
$Nc$ is the number of contacts}
\label{nmnc.ladder}
\end{figure}

The numerical solution  for $\frac{1}{N}\ln (N_{M})$ as a function of $c$, 
with the trivial end points $(0,0)$ and $(0.5,0)$,  are presented 
in Fig. \ref{nmnc.ladder}.

The final property of walks and maps on a ladder that we calculate deals with
the degeneracy of a  map with $Nc$ contacts. Denote by $G=e^{Ng}$  the 
number of walks that have the same map and by $H(g,c)$ 
the number of maps of $Nc$ contacts and  this value of the degeneracy. 
The quantity $H(g,c)$
is given by
\begin{equation}
H(g,c)=
\int_0^1  \pi dP(m) \delta \left[\sum_mP(m)\ln D(m)-g \right]
\delta\left[\sum_mmP(m)-1\right]
 \delta\left[\sum_mP(m)-c\right]
e^{-N \left[ -\sum_m P(m)\ln P(m)-c\ln c\right]}
\end{equation}
\noindent and the saddle point equations for $\{P(m)\}$ are 
\begin{eqnarray}
P(m)=P(2) D(m)\left[ {P(2)P(8) \over P(5)^2}\right]^{[\ln D(m)
/\ln 2]}
\left( {P(5) \over P(2)}\right)^{(m-2)/3}
\end{eqnarray}

\begin{figure}[h]
\centerline{\psfig{figure=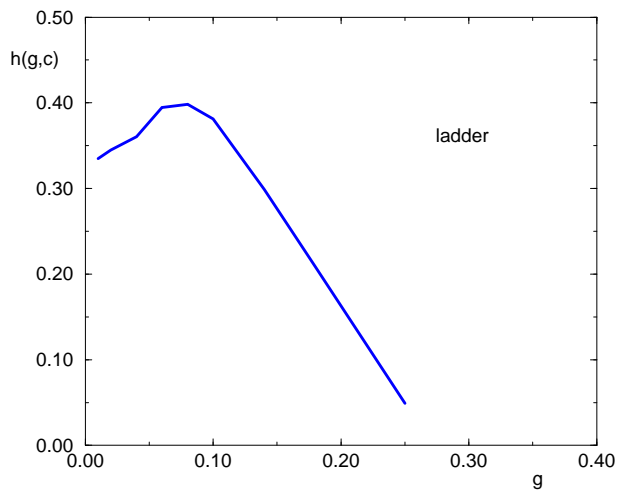,height=7.5cm,angle=0}}
\caption {Histogram of 
$h(g,c)=\ln H(g,c)/N$ versus  $g$, for $c=0.2$ on a ladder.}
\label{hg.ladder}
\end{figure}

\noindent The three unknown fractions 
$P(2),~P(5),~P(8)$ are determined through
the three global constraints 
\[
c=\sum_m P(m),\qquad 1=\sum_m P(m)m, \qquad 
g=\sum_m P(m)\ln D(m)
\]
A typical result for $c=0.2$ are presented in Fig. \ref{hg.ladder}. 

\begin{figure}[h]
\centerline{\psfig{figure=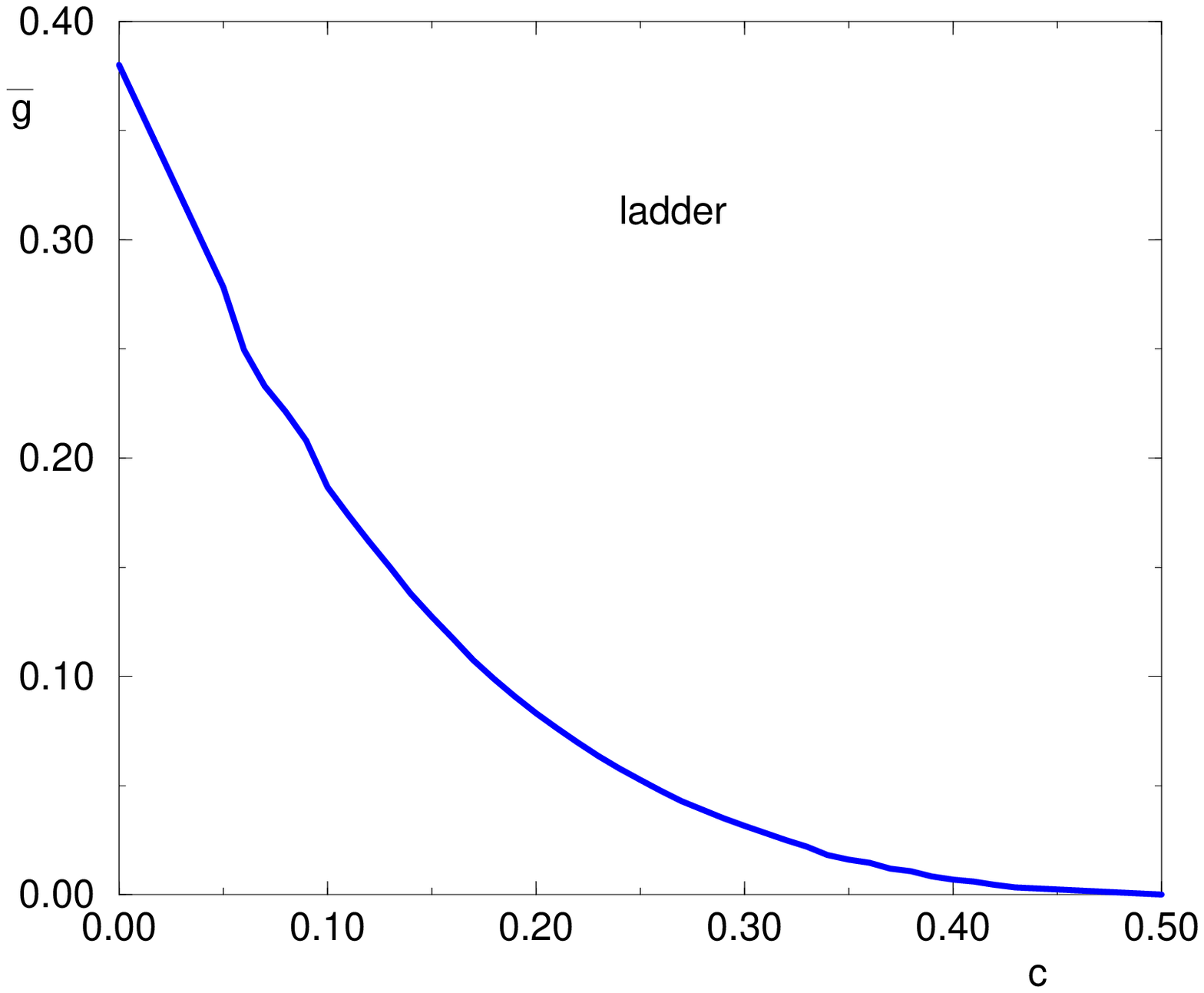,height=7.5cm,angle=0}}
\caption {Plot of $\overline{g}$ versus $c$ on a ladder.
}
\label{hag.ladder}
\end{figure}
\section{Semidirected Restricted Walks}
A related problem is that of semidirected restricted walks (SRW) 
on a square lattice. These walks are defined as follows:
all horizontal steps are directed - in the $+x$ direction. 
Vertical steps are restricted so that 
the number of consecutive vertical steps never exceeds
$ k$. The k=1 case is already a superset of  
walks on a ladder. 

The number of SRWs can be 
computed as follows.

Denote  
the total number of walks by $A(N)$. As before, $A_h(N)$ of these walks  
end with a horizontal step  and $A_v(N)$ walks end with a
vertical step. 
\begin{eqnarray*}
 A(N) = A_h(N)+ A_v(N)\\
\end{eqnarray*}
$A_v(N)$ can be further classified into k classes. ${A^i}_{v}(N)$ corresponds
to walks that end with exactly $i$ vertical steps. 
\begin{eqnarray*}
 A_v(N) = {A^1}_v(N)+{A^2}_v(N)+ \cdots + {A^k}_v(N) \\
 {A^i}_v(N)= {A^{(i-1)}}_v(N-1)\\ 
 {A^1}_v(N)= 2 A_h(N-1)\\
\end{eqnarray*}

A little algebra gives the following recursive relation

\begin{eqnarray*}
A(N) = A(N-1) + 2 \left( A(N-2)+ \cdots + A(N-k-1) \right)\\
\end{eqnarray*}

So the connective constant ( of exponential growth) is given by the root 
of the following polynomial equation: 

\begin{eqnarray*}
 (y-1) y^{k} = 2 \frac{1-y^{k}}{1-y}\\
\end{eqnarray*}
For $k=1$ this reduces to $(y-1)y=2$, i.e. $y=2$, whereas 
in the  $k \rightarrow \infty$ limit it simplifies to $(y-1)^2=2$ so that
the connective constant increases to $y=1+\sqrt{2} \approx 2.42$.
 
Computing the number of contact matrices for a general k seems slightly more 
tedious, but it is possible to do it explicitly for $k=1$.
We denote $B(N)$ by the number of distinct maps of size $N$. It can be 
classified into maps with either one contact or no contact  in the first 
row. The number of the former is $B_0(N)$ and the latter $B_1(N)$.

\begin{eqnarray*}
B(N) = B_0(N) +B_1(N) \\
B_0(N) = B(N-1) \\
B_1(N) = B(N-4) + B_1(N-2)\\
\end{eqnarray*}
 
A little algebra leads to the following recursive relation: 

\begin{eqnarray*}
B(N) = B(N-1)+B(N-2) - B(N-3)+B(N-4)\\
\end{eqnarray*}

which, in turn, leads to the following polynomial equation:

\begin{eqnarray*}
q^4 -q^3 -q^2 +q -1 = 0 \\
\end{eqnarray*}
The  root, $q \approx 1.51$,
corresponding to the growth factor for the maps, is  slightly higher 
than that of the ladder $(\approx 1.44)$. We have not found a simple way to 
compute $B(N)$ for general $k$ values. 
\section{Summary}
Contact maps are a compact and useful representation of a protein's
structure. Contact maps are used for screening candidate structures
from a database. More recently attempts were made to use them
to fold proteins, i.e. determine the map of a protein of known
sequence by minimizing some energy function.

In order to have a handle on the work involved in searching the
subspace of physical maps, it is important to know various statistics.
For example, how the number of physical maps increases with the 
protein's length, the dependence of various properties on the
number of contacts, etc. In this paper we studied these issues on
several lattices; for an essentially one-dimensional ladder the results
were obtained analytically and in two dimensions we studied the square
and triangular lattices	by exact enumeration and sampling. In addition
we provide exact bounds on the number of distinct physical maps, valid
in any dimension.

Our main findings can be summarized as follows:

\begin{itemize}
\item
The number of physical contact maps scales exponentially
with the length $N$ of the walk. 
\item
The number of contact maps (and of walks as well) is a non-monotonic 
function of the number of contacts.
\item
The average degeneracy of contact maps that have $N_C$ contacts 
decreases as $N_C$ increases.
\item
Contact maps corresponding to very compact walks 
(i.e. highest $N_C$) have low degeneracy.
The ground state of a protein is most likely
to be found among these maps.
\end{itemize}

{\bf Acknowledgments}

This work was initiated during a visit of ED to Rutgers, supported by
DIMACS.
BS and JL thank J. Kahn and O. Penrose for very helpful comments and 
discussions. BS and JL wish to acknowledge research 
supported in part by NSF grant  and DIMACS.
The work of MV and ED was partially supported by a grant from 
GIF (German Israel Science Foundation), the Israeli Ministry of Science
and the Minerva Foundation.


\end{document}